# Dynamical stability, Vibrational and optical properties of anti-perovskite $A_3BX$ (Ti$_3$TlN, Ni$_3$SnN and Co$_3$AlC) phases: a first principles study

K. Das[1], M. A. Ali[1], M. M. Hossain[1], S. H. Naqib[2], A. K. M. A. Islam[2,3] and M. M. Uddin[1*]

[1]Department of Physics, Chittagong University of Engineering and Technology, Chattogram-4349, Bangladesh

[2]Department of Physics, University of Rajshahi, Rajshahi 6205, Bangladesh

[3]Department of Electrical and Electronic Engineering, International Islamic University Chittagong, Kumira, Chittagong, 4318, Bangladesh

**Abstract:**

We have investigated various physical properties including phonon dispersion, thermodynamic parameters, optical constants, Fermi surface, Mulliken bond population, theoretical Vickers hardness and damage tolerance of anti-perovskite $A_3BX$ phases for the first time by employing density functional theory (DFT) methodology based on first principles method. Initially we assessed nine $A_3BX$ phases in total and found that only three phases (Ti$_3$TlN, Ni$_3$SnN and Co$_3$AlC) are mechanically and dynamically stable based on analysis of computed elastic constants and phonon dispersion along with phonon density of states. We revisited the structural, elastic and electronic properties of the compounds to judge the reliability of our calculations. Absence of band gap at the Fermi level characterizes the phases under consideration as metallic in nature. The values of Pugh ratio, Poisson's ratio and Cauchy factor have predicted the ductile nature associated with strong metallic bonding in these compounds. High temperature feasibility study of the phases has also been performed using the thermodynamic properties, such as the free energy, enthalpy, entropy, heat capacity and Debye temperature. The Vickers hardness of the compounds are estimated to be ~ 4 GPa which is comparable to many well-known *MAX* phases, indicating their reasonable hardness and easily machinable nature. The static refractive index $n(0)$ has been found around ~ 8.0 for the phases under study that appeals as potential candidate to design optoelectronics appliances. The reflectivity is found above 44% for the Ti$_3$TlN compound in the energy range of 0-14.8 eV demonstrating that this material holds significant promise as a coating agent to avoid solar heating.

*Corresponding authors: mohi@cuet.ac.bd

1. **Introduction**

One of the biggest challenges in designing the structural materials is that it should have high stiffness (i.e., should be reasonably hard materials) associated with damage tolerant behavior. Ductility assessment is one of the ways to define a material as damage tolerant. Usually, ceramics possess excellent high temperature mechanical properties, high elastic moduli and oxidation and corrosion resistance but they are brittle in nature [1,2]. Metals, on the other hand, are ductile but soft that limits their use as strong structural materials [3]. In the recent years, attempts have been made to combine the properties of ceramics and metals that should have high elastic moduli and should be ductile in nature at the same time. For example, some binary carbides and nitrides are known to have high elastic moduli and brittleness [4-8]. Their damage tolerant behavior can be improved by incorporating a weak layer to form laminated structure with alternative strong and weak stacking layers. The so called *MAX* phase nanolaminates are the famous examples of layered structure in which non-directional metallic bonds are formed in addition to strong directional covalent bonds [9-16]. The *MAX* phases are defined with general formula M$_{n+1}$AX$_n$ with n = 1-3; M = early transition metal; A= A-group elements from group 12-16 in the periodic table and X= C and/or N) [17]. Besides the *MAX* phases, excellent machinability and damage tolerant behavior are also reported in Zr$_3$AlN, Hf$_3$AlN, Cr$_2$AlB$_2$, Zr$_2$Al$_3$C$_5$ [18-20]. In general, often ductility is improved but the stiffness of these materials is decreased due to the existence of the weak layers synergistically [18-20]. This limitation could be overcome by approaching new strategy where unlike *MAX* phases, there is a metallic structural unit in which the strong covalent bonds exist within the $A_3BX$ ($A$ and $B$ are different metals and $X$ is C or N) formula. Recently, Zhang et al. [21] reported a family of anti-perovskite $A_3BX$ structure with high stiffness and excellent damage tolerant properties. Within this structure $A_3B$ acts as metallic box that contribute to achieve ductility and octahedra $XA_3$ are in corners of the structure that contribute to strong covalent bonding leading to high stiffness. They have studied 126 $A_3BX$ phases based on the mechanical properties and predicted six hard and four soft compounds which have shown excellent damage tolerant with highly stiff ceramic nature.

The detail theoretical understanding of physical properties of the materials is essential to select compounds to be used in various technological applications. Dynamical stability check is also required before approval of the materials for practical application under external pressure and temperature conditions. Furthermore, the materials' response to high temperature and pressure can be understood by the study of their thermodynamic behavior that is considered as the root criteria for many industrial applications [22]. Additionally, investigations of optical properties such as reflectivity, absorption coefficient and refractive index of the materials are crucial to predict the suitability of those to be used as solar reflector, solar absorber, coating agent to reduce solar heating and possible optoelectronic device applications [23, 24].

We have investigated nine promising damage tolerant ceramics ($Ti_3AlN$, $Mn_3CuN$, $Ti_3TlN$ and $Ni_3SnN$ that are predicted to be soft and $Mn_3NiN$, $Mn_3GaN$, $Mn_3SnC$, $Cr_3SnN$ and $Co_3AlC$ that are supposed to have high elastic moduli) from the predicted anti-perovskite $A_3BX$ family [21]. Among those, only three, namely, $Ti_3TlN$, $Ni_3SnN$ and $Co_3AlC$ are found to show dynamical stability. Therefore, we focused to perform a detailed study of these three dynamically stable $Ti_3TlN$, $Ni_3SnN$ and $Co_3AlC$ compounds. We have explored the mechanical anisotropy, Fermi surface topology, Mulliken bond population analysis, Vickers hardness, phonon dispersion, thermodynamic and optical properties for the first time for these three anti-perovskite compounds. We have also revisited the elastic and electronic properties of the Ti$_3$TlN, Ni$_3$SnN and Co$_3$AlC to check the consistency of our calculations as well as to validate prior investigations.

## 2. Computational methodology

The ground state electronic structure with equilibrium crystal structure have been investigated using the CAmbridge Serial Total Energy Package (CASTEP) code [25] in which the pseudo-potential plane-wave (PP-PW) total energy calculation approach based on the density functional theory (DFT) [26] are employed. The interactions of electrons with ion cores are characterized by pseudo-potentials of Vanderbilt-type ultrasoft formulation [27]. The electronic exchange correlation energy is modeled under the Generalized Gradient Approximation (GGA) method in conjunction with the Perdew-Burke-Ernzerhof (PBE) formalism [28]. On the other hand, the phonon contributions to the structural stability and thermodynamic properties of the compounds have been evaluated using the norm-conserving pseudo-potential. The plane wave cut-off energy is set to 500 eV for all the calculations to ensure convergence. The *k*-points (6×6×6) sampling integration is fixed to ultrafine quality over the first Brillouin zone for the crystal structure optimization by using the Monkhorst-Pack scheme [29]. The structural parameters including lattice constants and internal atomic coordinates and consequent charge density of cubic Ti$_3$TlN, Ni$_3$SnN, and Co$_3$AlC are evaluated by employing the Broyden–Fletcher–Goldfarb-Shanno (BFGS) [30] minimization technique. The periodic boundary conditions are also used in the total energy of each cell. The tolerances for geometric optimization are selected as follows: total energy within $5.0\times10^{-6}$ eV/atom, self-consistent field within $5.0\times10^{-7}$ eV/atom, maximum ionic Hellmann–Feynman force within 0.01 eV/Å, maximum ionic displacement within $5.0\times10^{-4}$ Å, and a maximum stress of 0.02 GPa. The elastic constant tensors, $C_{ij}$ have been calculated by the 'stress-strain' method in-built in the CASTEP program that is also used to evaluate bulk modulus $B$ and shear modulus $G$ of the compounds of interest.

## 3. Results and discussion

### 3.1 *Structural properties*

The structure of *A₃BX* phases crystallizes with the space group *Pm-3m* (221) belonging to the cubic system [31]. The unit cell structure of *A₃BX* [Ti₃TlN (TTN), Ni₃SnN (NSN), and Co₃AlC (CAC)] compounds is illustrated in Fig. 1. There are five atoms in the unit cell with the Wyckoff positions of the *A* atoms at (0, 0.5, 0.5), *B* atoms at (0, 0, 0) and *X* atoms at (0.5, 0.5, 0.5), respectively [32], as displayed in Fig. 1. The optimization of the equilibrium crystal structures of TTN, NSN and CAC are achieved by minimizing the total energy and the structural parameters of the compounds are given in Table 1. Our calculated results are compared with available theoretical and experimental results as listed in Table 1, which affirms the reliability and accuracy of our calculations. The obtained lattice constants *a* and cell volume of the compounds TTN, NSN and CAC are deviated less than 1.7% from available reported values (Table 1) and offer a very good validation of the earlier theoretical estimates [21, 33-35].

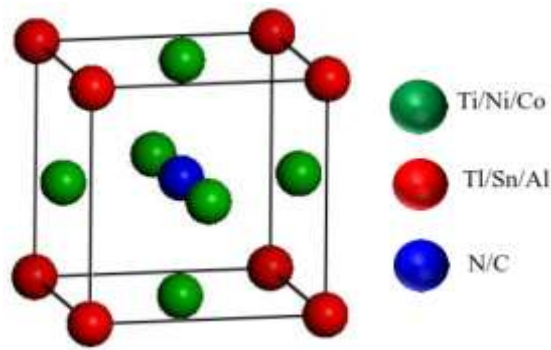

**Fig. 1:** The schematic unit cell structure of the *A₃BX* (Ti₃TlN, Ni₃SnN, and Co₃AlC) compounds.

**Table 1.** Different structural parameters of the $A_3BX$ (Ti$_3$TlN, Ni$_3$SnN and Co$_3$AlC) compounds [lattice parameters $a$ (Å) and unit cell volume V (Å$^3$)].

| Phases | A | V | Deviation | | Ref. |
|---|---|---|---|---|---|
| | | | $a$ | V | |
| | 4.212 | 74.720 | | | This work |
| Ti$_3$TlN | 4.214 | 74.833 | 0.05% | 0.15% | Theo. ICSD [33] |
| | 4.159 | | 0.2% | | Theo. [21] |
| | 3.926 | 60.536 | 0.4% | 1.3% | This work |
| Ni$_3$SnN | 3.909 | 59.709 | | | Theo. ICSD [34] |
| | 3.850 | | 1.5% | | Theo. [21] |
| | 3.744 | 52.475 | 1.1% | | This work |
| Co$_3$AlC | 3.700 | | | | [35] |
| | 3.680 | | 1.7% | | Theo [21] |

*3.2 Mechanical properties*

The study of mechanical properties such as stability, elastic moduli, stiffness, brittleness, ductility, and elastic anisotropy of a material provides with fundamental information required for selected engineering applications. The linear finite strain-stress method within the CASTEP code [36] has been used to calculate the elastic constants. The single crystal elastic constants $C_{ij}$ and various polycrystalline elastic moduli of the compounds are represented in Table 2 along with available reported results. It is seen that all the compounds under consideration show very large tensile related $C_{11}$ and low values of $C_{12}$ and $C_{44}$ which are related to shear. The compounds satisfy the Born stability [37] criteria expressed as: $C_{11} > 0$, $C_{11}$-$C_{12} > 0$, $C_{44} > 0$, $(C_{11} + 2C_{12}) > 0$ which together endorse the mechanically stability of the crystals.

**Table 2.** The elastic constants, $C_{ij}$ (GPa) and various moduli [bulk B (GPa), shear, G (GPa), Young's, Y (GPa)], Pugh ratio, G/B, Poisson ratio, ν, and Cauchy Pressure of damage tolerant and high stiffness $Ti_3TlN$, $Ni_3SnN$ and $Co_3AlC$ compounds.

| Phases | $C_{11}$ | $C_{12}$ | $C_{44}$ | A | B | G | Y | G/B | ν | Cauchy Pressure | Ref |
|---|---|---|---|---|---|---|---|---|---|---|---|
| $Ti_3TlN$ | 196 | 131 | 52 | 1.6 | 153 | 43 | 118 | 0.276 | 0.40 | 79 | This work |
|  | 228 | 135 | 44 | 0.95 | 166 | 45 | 124 | 0.27 | 0.38 | 91 | Theo.[21] |
| $Ni_3SnN$ | 266 | 132 | 41 | 0.61 | 177 | 50 | 137 | 0.282 | 0.37 | 91 | This work |
|  | 305 | 141 | 45 | 0.55 | 196 | 58 | 157 | 0.29 | 0.37 | 96 | Theo.[21] |
| $Co_3AlC$ | 451 | 119 | 86 | 0.52 | 230 | 112 | 290 | 0.487 | 0.29 | 38 | This work |
|  | 498.1 | 133.1 | 96.4 | 0.53 | 255 | 125 | 322 | 0.49 | 0.29 | 36.65 | Theo.[21] |

The compounds TTN and NSN show exceptionally low $C_{44}$ value that are related to shear deformation and damage tolerant behavior, while higher $C_{44}$ observed for the CAC results in higher shear modulus (G) as shown in Table 2. The shear anisotropy factor (Zener ratio) A, is defined by [38]

$$A = \frac{G}{Y/[2(1+\nu)]} = \frac{2C_{44}}{C_{11} - C_{12}}$$

This index designates the anisotropic nature of the compounds with the possibility of appearance of microcracks. The value of A is unity for completely isotropic material and any deviation from unity (smaller or greater) denotes the degree of anisotropy. The computed values of A (Table 2) indicate anisotropic nature of the compounds under study. Moreover, our calculated stiffness constants ($C_{ij}$) for the compounds are consistent with previously reported results [21].

In most of the cases, contour plots (3D) and their two dimensional (2D) views are very effective to visualize the anisotropic mechanical behavior of materials and for better understanding of elastic responses as well. We have calculated the Young's modulus (Y), shear modulus (G) and Poisson's ratio and plotted them in 3D contour plots and in their 2D projections [39]. The plots are almost similar and identical for the compounds under investigation. Therefore, only plots for the TTN compound have been displayed in Fig. 2 as representative. It is clear that the values of Y, G and ν are deviated from the spherical shape in 3D and from circles in 2D, indicating the anisotropic nature of the compound. The ratios of maximum and minimum values of Y, G and ν define the anisotropy factor (AF) in this particular scheme [39] that signifies the degree of anisotropy and are represented in Table 3. It is seen from the Table 3 that the AF is almost equal for the TTN and NSN compounds, however it is a bit higher for the CAC compound due to its higher elastic moduli compared to the other two members.

The single crystal elastic constants, $C_{ij}$ and respective compliance tensors $S_{ij}$ ($S_{ij} = C_{ij}^{-1}$) have been used to evaluate the various elastic moduli ($B$, $G$, $Y$, and $v$) using the following formulae: $Y = \frac{9B_H G_H}{(3B_H + G_H)}$ and $v = \frac{(3B_H - Y_H)}{6B_H}$ [40,41]. These values are given in Table 2 and all polycrystalline elastic moduli (B, G, Y) are defined by the Voigt-Reuss-Hill (VRH) [42-44] equations as follows: $B_V = \frac{C_{11} + 2C_{12}}{3}$ and $G_V = \frac{C_{11} - C_{12} + 3C_{44}}{5}$, $B_R = \frac{1}{3s_{11} + 6s_{12}}$ and $G_R = \frac{5}{4(s_{11} - s_{12}) + 3s_{44}}$, $B_H = \frac{B_R + B_V}{2}$ and $G_H = \frac{G_R + G_V}{2}$. The compounds retain significant overall bonding strength of the atoms as indicated by the calculated values of $B$ [45]. The hardness of the compounds are in general measured by the values of $B$ (G) and are found to be 153 (43) GPa, 177 (50) GPa and 230 (112) GPa for the compounds TTN, NSN and CAC, respectively. It can be anticipated that the CAC compound would be significantly harder than TTN and NSN due to higher values of both the elastic moduli. The Young's modulus (Y) of CAC is also much higher than that of TTN and NSN indicating that CAC is also stiffer than the others. The three factors, Paugh ratio (G/B) [46], Poisson's ratio $v$ and Cauchy pressure are sensible criteria to classify materials either as ductile or as brittle. It is seen from Table 2 that the values of *G*/*B* are found to be less than 0.57 and the Poisson's ratios 0.40, 0.37 and 0.29 for the TTN, NSN and CAC compounds, respectively, are greater than the value given by Frantsevich's criterion (0.26) [47]. These indicators demonstrate the ductile nature of the compounds under consideration and suggest that there should be metallic bonds in the three anti-perovskites.

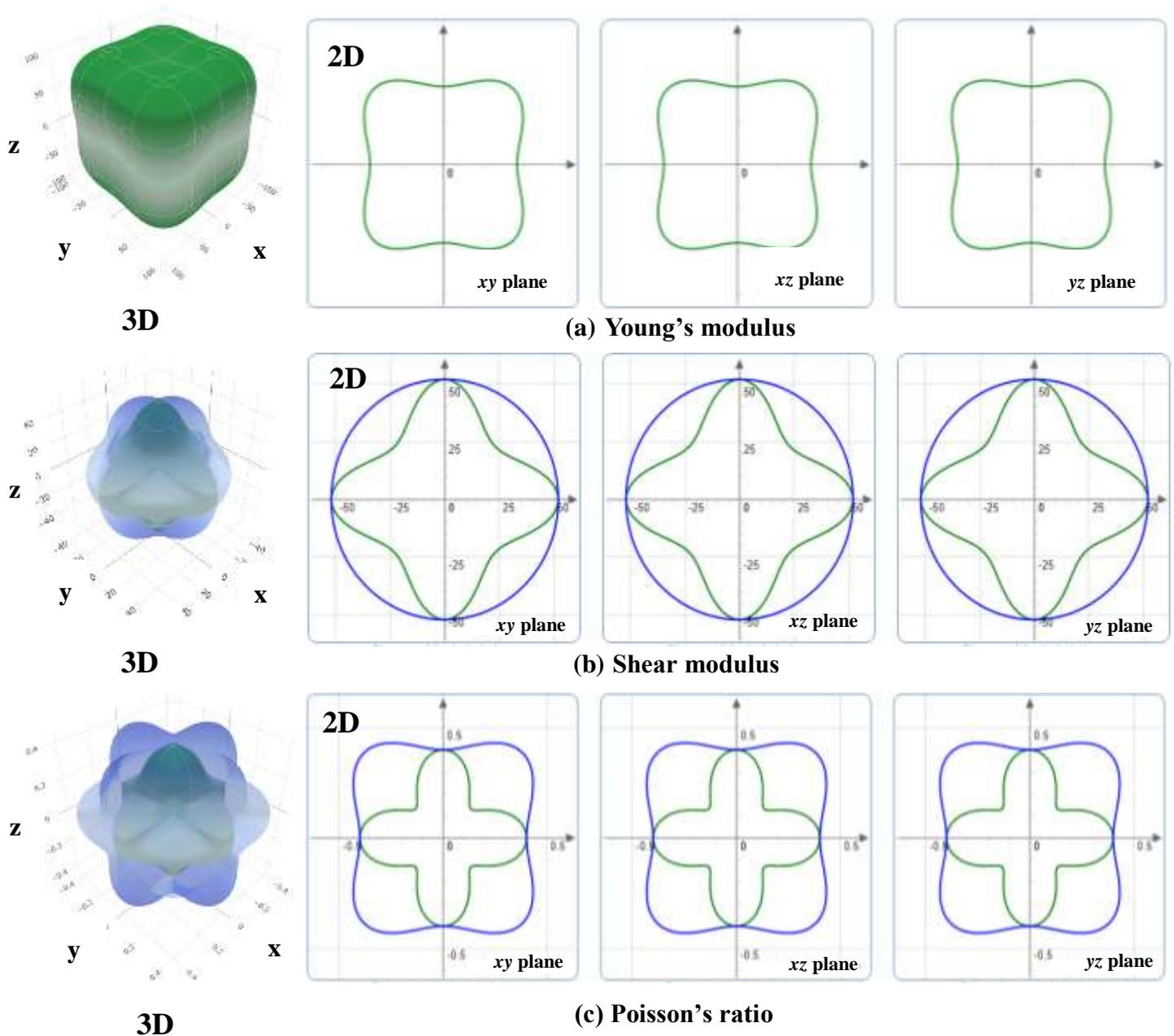

**Fig. 2:** Contour (3D) and two dimensional (2D) plots of (a) Young's modulus (*Y*) (b) shear modulus (*G*) and (c) Poisson's ratio (υ) for the anti-perovskite Ti$_3$TlN phase.

The Cauchy pressure ($C_{12} - C_{44}$) [48] can also be used to examine the bonding nature of solids. A negative value indicates that the compounds is dominated by strong directional bonding (covalent) with brittle nature while a positive value endorses the characteristics related to metallic bondings with quasi-ductile nature. Moreover, the high bulk modulus with low shear modulus of the compounds clearly demonstrates the damage tolerant, easily machinable, quasi-ductile and stiff nature of the materials.

**Table 3**: The minimum and maximum limit of Young's modulus, $Y$ (GPa), shear modulus, $G$ (GPa) and Poisson's ratio, $\upsilon$ of the anti-perovskite $A_3BX$ phases.

| Compound | Young's modulus | | | Shear modulus | | | Poisson's ratio | | |
|---|---|---|---|---|---|---|---|---|---|
| | $Y_{min}$ | $Y_{max}$ | $Y_{max}/Y_{min}$ | $G_{min}$ | $G_{max}$ | $G_{max}/G_{min}$ | $\upsilon_{min}$ | $\upsilon_{max}$ | $\upsilon_{max}/\upsilon_{min}$ |
| Ti$_3$TlN | 91.04 | 140.09 | 1.54 | 32.5 | 52 | 1.6 | 0.19 | 0.54 | 2.84 |
| Ni$_3$SnN | 114.17 | 178.44 | 1.56 | 41 | 67 | 1.6 | 0.23 | 0.53 | 2.30 |
| Co$_3$AlC | 229.37 | 401.31 | 1.75 | 86 | 166 | 1.93 | 0.13 | 0.49 | 3.77 |

### 3.3 *Electronic properties*

#### 3.3.1 *Band structure and density of states (DOS)*

The electrical properties of solids can be understood from the electronic band structure of the compounds. The energy band structures of $A_3BX$ phases [Ti$_3$TlN (TTN), Ni$_3$SnN (NSN), and Co$_3$AlC (CAC)] along high-symmetry $k$ points in the first Brillouin zone have been calculated using the equilibrium lattice parameters and are shown in Fig. 3(a, b, c), respectively. The conduction bands and valence bands at the Fermi level (solid black line) are significantly overlapping which endorses the metallic nature of the materials. A complicated mixed band character including the quasi-flat band for the TTN compound with a series of highly dispersive bands intersecting the $E_F$ have been observed at and in the vicinity of the

Fermi level. It is seen from the band structure that the electrical conductivity along the *c* direction is lower than that in the *ab* plane, since the energy dispersion along the *c* direction is lower.

Total and partial density of states (TDOS and PDOS, respectively) of the phases assist to fully understand the detailed features of band structure. The calculated TDOS and PDOS of the $A_3BX$ phases (Ti$_3$TlN (TTN), Ni$_3$SnN (NSN), and Co$_3$AlC (CAC)) have been illustrated in Figs. 3 (d-f).

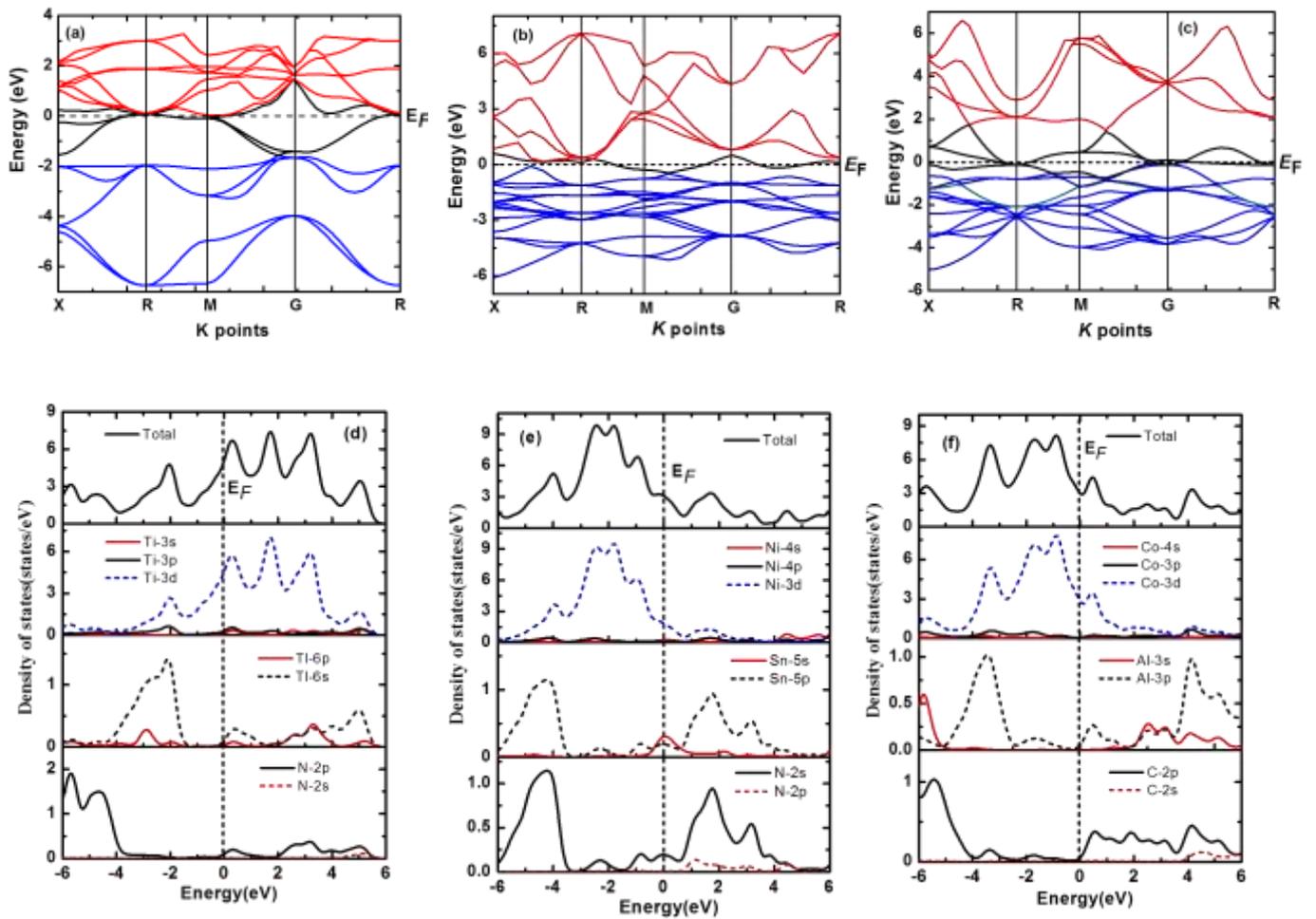

**Fig. 3.** Band structures of the $A_3BX$ phases. (a) Ti$_3$TlN, (b) Ni$_3$SnN, and (c) Co$_3$AlC. The total and partial density of states (DOS) of (d) Ti$_3$TlN, (e) Ni$_3$SnN, and (f) Co$_3$AlC.

The vertical black dashed line specifies the Fermi level, $E_F$. The valence and conduction bands are overlapped consequently no band gap can be observed at the Fermi level which implies that the compounds are metallic in nature. The value of TDOS at the $E_F$ is found to be 5.02, 3.07 and 2.97 states per eV for the phases TTN, NSN and CAC, respectively. The electrons in the Ti-3$d$, Ni-3$d$ and Co-3$d$ orbitals contribute strongly in the electronic conduction properties of the TTN, NSN and CAC phases, respectively.

Basically, the valence band can be separated into three sub-bands. The lower ones from -6 to -4.4, -6 to -3.5, -6 to -2.7 eV, the middle ones from -4.4 to -1.14, -3.5 to -1.2, -2.7 to -0.6 eV and the top sub-bands crossing the $E_F$ for TTN, NSN and CAC, respectively. Respective contributions come mostly from N-2$p$ and Tl-6$s$, N-2$s$ and Sn-5$s$, C-2$p$ and Al-3$p$ and Ti-3$d$, Ni-3$d$ and Co-3$d$ electronic states for the phases TTN, NSN and CAC, respectively as displayed in Figs. 3 (d-f). At the $E_F$, the Ti-3$s$, 3$p$, Tl-6$p$ and Ni-4$s$, Sn-5$p$ are moderately contributing to the TDOS of TTN and NSN phases. The TDOS values for TTN and NSN are larger than that of the CAC phase (2.97 states per eV) where Co-4$s$ 3$p$ and Al-3$s$ electronic states do not contribute significantly to the TDOS at the Fermi level. It suggests that stiffness of the CAC phase is higher than that of TTN and NSN owing to the lower TDOS at $E_F$ as suggested by Liao et al. [49]. Three bands are mixed together strongly and crossed the $E_F$ for the phases TTN and CAC, however only one band crossed the Fermi level for NSN, consequently higher values of Vickers hardness are expected for TTN and CAC compared to that of the NSN phase. The electronic charge density distribution mapping provides with more information to understand the bonding contributions of electrons in the different bands. This feature will be discussed in the next section.

*3.3.2 Charge density distribution mapping and Fermi surface*

The electronic charge density distribution mapping in the contour form (in the units of e/Å$^3$) in the (101) crystallographic plane of Ti$_3$TlN, Ni$_3$SnN and Co$_3$AlC phases have been calculated and are shown in Figs. 4(a-c). The high and low values of electronic charge density in the adjacent map are indicated by the colors red and blue, respectively. The positive value (accumulation of charges) indicates the formation of covalent bonds between two atoms while the negative value (depletion regions) arises from charge transfer from specific atom facilitating the formation of ionic bonding [50, 51]. The strong charge accumulation is observed between Co, C and Ni, Sn in the CAC and NSN phases, respectively, as shown in Fig. 4 (b, c). Accordingly strong covalent bondings are formed between Co-C, and Ni-Sn atoms. Moreover, the less charge accumulation between Co, Al and Ni, N results in partial metallic bonds for Co-Al and Ni-N in the phases CAC and NSN, respectively. The charge accumulation is absent between Ti and Tl in the TTN phase (Fig. 4 (a)) indicating strong metallic bond arises between these atoms in the *A$_3$B* box of the crystal structure. The covalent-ionic bonding between N and C has been inferred in the phases which are comparatively weaker (Fig. 4 (a-c)). The observed metallic bonds strengthen the damage tolerant properties of the phases under investigation. It should be noted that the covalent-ionic bonds (*A-X*) are liable for high stiffness of the materials while metallic or metallic-covalent bonds between *A-B* governs the shear deformation (damage tolerant) related features.

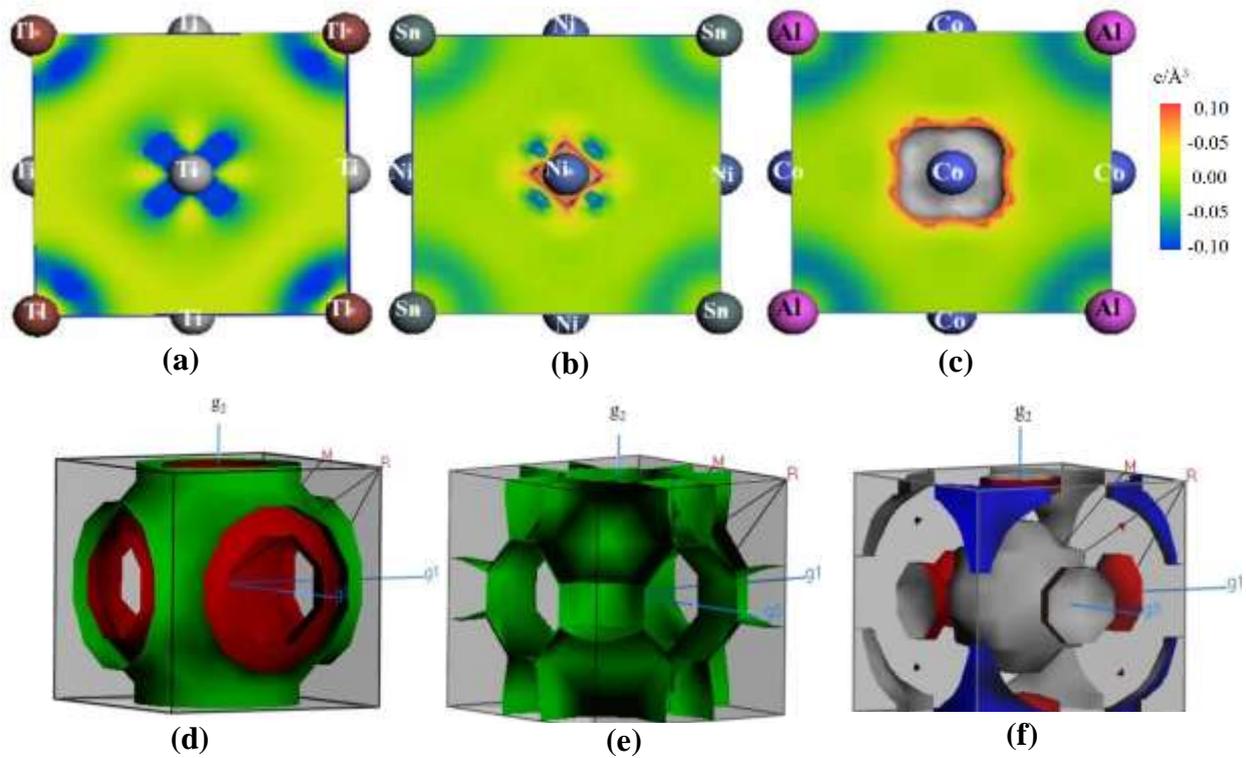

**Fig. 4.** Electronic charge density mappings for (a) $Ti_3TlN$, (b) $Ni_3SnN$, and (c) $Co_3AlC$. The Fermi surfaces of (d) $Ti_3TlN$, (e) $Ni_3SnN$, and (f) $Co_3AlC$.

The Fermi surface topology of the $A_3BX$ phases ($Ti_3TlN$, $Ni_3SnN$, and $Co_3AlC$) have been evaluated at zero pressure in the equilibrium structure and are depicted in Fig. 4(d-f). Both electron and hole-like sheets have been observed. The Fermi surface shapes of the compounds are different in nature and easily distinguishable for all the compounds. Three Fermi surface sheets are identified for three crossing band at the $E_F$ for two phases of TTN and CAC while one sheet is detected for the NSN phase, as indicated by the band structure calculations. However, the third sheet (centre sheet) of the TTN phase is almost attached with the second sheet (red color) in Fig. 4d that is somewhat difficult to separate due to very low dispersion as shown in PDOS profile in Fig 3d. For the TTN, first (green) and second (red) sheets have hexagonal cross sections (hollow cylindrical shape) along the Γ-M and Γ-R directions of the Brillouin zone, respectively (Fig. 4d). For the CAC compound, first (centre, ash color) sheet is spherical in shape along the Γ-M direction, second sheet is octagon shaped in the middle of the cubic plane along the Γ-R direction and third (outer most) sheet is ribbon-like tubes located in each corners in the Γ-R direction (Fig. 4f). In case of NSN, the sheet is hollow cylindrical shaped along Γ-M and Γ-R directions with six holes at cubic structural planes. It is noteworthy that for the NSN phase, several bands are very close to the $E_F$ (Fig. 3b) that is well reflected in the Fermi surface topology (Fig. 4e) and it seems another sheet might be attached with crossing sheet along Γ-R direction as shown in Fig. 4e. The Fermi surfaces of the TTN and CAC compounds are formed due to the highly dispersive Ti-$3d$, Tl-$6s$, N-$2p$ and Co-$3d$, Al-$3p$, C-$2p$ electronic states, respectively. For the NSN compound, it is due to lowly dispersive Ni-$3d$, Sn-$5s$ and N-$2s$ orbitals.

3.3.3 *Mulliken populations study*

Distribution of electron density in different bonds can be explained by the Mulliken bond population analysis. The bond-overlap population (BOP) analysis provides with quantitative measures of bonding and anti-bonding strengths [50-52]. The values of BOP are interpreted as follows: i) *zero*: no significant interaction between electronic populations of the two atoms involved and it is ignored to calculate the hardness of the materials, ii) *positive*: the neighbor atoms are bonded, iii) *negative*: they are anti-bonded. A high value of BOP indicates high degree of covalency of the bonds. The effective valance charge (EVC) is defined as the difference between formal ionic charge and Mulliken charge within a crystal. The value of EVC is used to guess the nature of the bonds; whether ionic or covalent. The EVC is zero (positive) means existence of ideal ionic bond (increasing level of covalency). It is seen that anti-bonding exists between Ti-Tl, Ni-Ni and Co-Co atoms in the TTN, NSN and CAC phases, respectively, since the bond overlap population is negative (Table 4). The bonds between Ti/Ti, N/Ni and Al/Co demonstrate higher degree of covalency than that of N/Ti, Ni/Sn and C/Co in the compounds as shown in Table 4. The bond Ti-Ti shows the highest level of covalency among the bonds in the compounds under study. Moreover, the overall bond strength is greater in the TTN and CAC than that in NSN. Hence, the hardness value is expected to be higher for those two phases. The charge transfer mechanism can also be understood using the atomic population analysis. For example, in TTN, charge transfers from Ti to Tl and N with the values of 0.78e and 0.75e, respectively. Similar features can be seen in other compounds.

**Table 4.** Mulliken atomic and band overlap populations of $Ti_3TlN$, $Ni_3SnN$ and $Co_3AlC$ compounds.

| Phases | Atoms | Mulliken atomic population | | | | | | Mulliken bond overlap population | | | |
|---|---|---|---|---|---|---|---|---|---|---|---|
| | | $s$ | $p$ | $d$ | Total | Charge (e) | EVC (e) | Bond | Bond number $n^\mu$ | Bond length $d^\mu$ (Å) | Bond overlap population $P^\mu$ |
| $Ti_3TlN$ | N | 1.67 | 4.07 | 0.00 | 5.75 | -0.75 | 3.75 | N-Ti | 3 | 2.06 | 0.61 |
| | Ti | 2.13 | 6.56 | 2.79 | 11.49 | 0.51 | 3.49 | Ti-Ti | 3 | 2.91 | 0.81 |
| | Tl | 1.35 | 2.38 | 10.04 | 13.78 | -0.78 | 3.78 | Ti-Tl | 3 | 2.91 | -2.40 |
| $Ni_3SnN$ | N | 1.62 | 4.03 | 0.00 | 5.65 | -0.65 | 3.65 | N-Ni | 3 | 1.90 | 0.63 |
| | Ni | 0.43 | 0.71 | 8.82 | 9.96 | 0.04 | 3.96 | Ni-Sn | 3 | 2.69 | 0.36 |
| | Sn | 0.99 | 2.49 | 0.00 | 3.48 | 0.52 | 4.48 | Ni-Ni | 3 | 2.69 | -0.50 |
| $Co_3AlC$ | C | 1.5 | 3.23 | 0.00 | 4.73 | -0.73 | 2.73 | C-Co | 3 | 1.83 | 0.57 |
| | Al | 0.94 | 1.88 | 0.00 | 2.82 | 0.18 | 2.82 | Al-Co | 3 | 2.59 | 0.63 |
| | Co | 0.39 | 0.55 | 7.88 | 8.82 | 0.18 | 3.82 | Co-Co | 3 | 2.59 | -1.12 |

**Table 5.** Calculated Mulliken populations [$\mu$-type bond $P^\mu$, bond length $d^\mu$, metallic population $P^{\mu'}$, bond volume $v_b^\mu$, Vickers hardness of $\mu$-type bond $H_V^\mu$ and $H_V$] of Ti$_3$TlN, Ni$_3$SnN and Co$_3$AlC.

| Phases | Bond | $d^\mu$ | $P^\mu$ | $P^{\mu'}$ | $v_b^\mu$ | $H_V^\mu$ | $H_V$ |
|---|---|---|---|---|---|---|---|
| Ti$_3$TlN | N-Ti | 2.06112 | 0.61 | 0.0614 | 6.505 | 17.906 | 3.60 |
|  | Ti-Tl | 2.91486 | 0.81 | 0.0614 | 24.90 | 2.6079 |  |
| Ni$_3$SnN | N-Ni | 1.90426 | 0.63 | 0.0284 | 5.270 | 27.896 | 3.58 |
|  | Ni-Sn | 2.69303 | 0.36 | 0.02824 | 20.17 | 1.641 |  |
| Co$_3$AlC | C-Co | 1.8356 | 0.57 | 0.1764 | 4.568 | 23.155 | 4.04 |
|  | Al-Co | 2.59593 | 0.63 | 0.1764 | 17.491 | 2.848 |  |

3.4 *Vickers Hardness*

The hardness provides a significant role for understanding mechanical behavior of the materials and assists one to select them for specific engineering applications. We have calculated the hardness of the studied phases using established formalism [53, 54] and presented those in Table 5.

$$H_V = \left[ \prod^\mu \left\{ 740(P^\mu - P^{\mu'})(v_b^\mu)^{-5/3} \right\}^{n^\mu} \right]^{1/\sum n^\mu}$$

where $P^\mu$ is the Mulliken population of the $\mu$-type bond, $P^{\mu'} = n_{free}/V$ is the metallic population, and $v_b^\mu$ is the bond volume of $\mu$-type bond. The Vickers hardness values are found to be 3.6, 3.58 and 4.04 GPa for the phases TTN, NSN and CAC, respectively. The Vickers hardness is highest for the CAC compound that is expected from its high elastic moduli and charge density mapping as well. The hardness values are comparable with those for many well-known *MAX* phase nanolaminates (in the range 2 - 8 MPa) such as for $Hf_2InC$ (3.45 MPa) and $Ta_2InC$ (4.12 MPa) [51].

*3.5 Phonon dispersion curve*

The materials' structural stability and vibrational contribution to the thermodynamic properties such as thermal expansion, Helmholtz free energy and heat capacity can be understood by the study of phonon dispersion curve (PDC) and phonon density of states (PHDOS). The PDC and PHDOS have been calculated of the phases TTN, NSN, and CAC along the high symmetry directions of the Brillouin zone (BZ) using the density functional perturbation theory (DFPT) based linear-response method [55-57]. The PDC and PDOS profiles are shown in Fig. 5 (a-f). The dependence of phonon frequency on wave vector ($k$) for the low frequency acoustic and high frequency optical modes with an energy gap at the edge of the BZ has been observed (Fig. 5 a,c,e) for all three anti-perovskites. The acoustic (lower) and optical (upper) branches exhibit a clear gap. The optical branches are located at the top of the PDC curves such as the transverse optical (TO) and the longitudinal optical (LO) branches. These branches play vital roles to realize the optical response of lattice vibrations in solids. The values of corresponding frequencies at the point M of TO and LO are estimated to be 16.47, 15.52 THz (TTN), 16.03, 14.37 THz (NSN) and 25.77, 23.01 THz (CAC), respectively.

To identify the bands and corresponding structures in the PHDOS of TTN, NSN, and CAC phases, we have shown them side by side for better understanding in Figs. 5 (b, d, f), respectively. It is observed that the prominent peaks arise due to the flatness of the bands of TO (LO) for the phase TTN (NSN and CAC) while non-flat bands results in weak peaks in the PHDOS. The calculated PDC of the phases reveals that the frequencies of phonon modes throughout the Brillouin zone for all wave vectors are positive with no imaginary (negative in the frequency scale) component, demonstrating dynamical stability of TTN, NSN and CAC

(Fig.5 a, c, e) compounds. This result is consistent with the mechanically stability of the phases that has been proven using stiffness constants as discussed in Section 3.2.

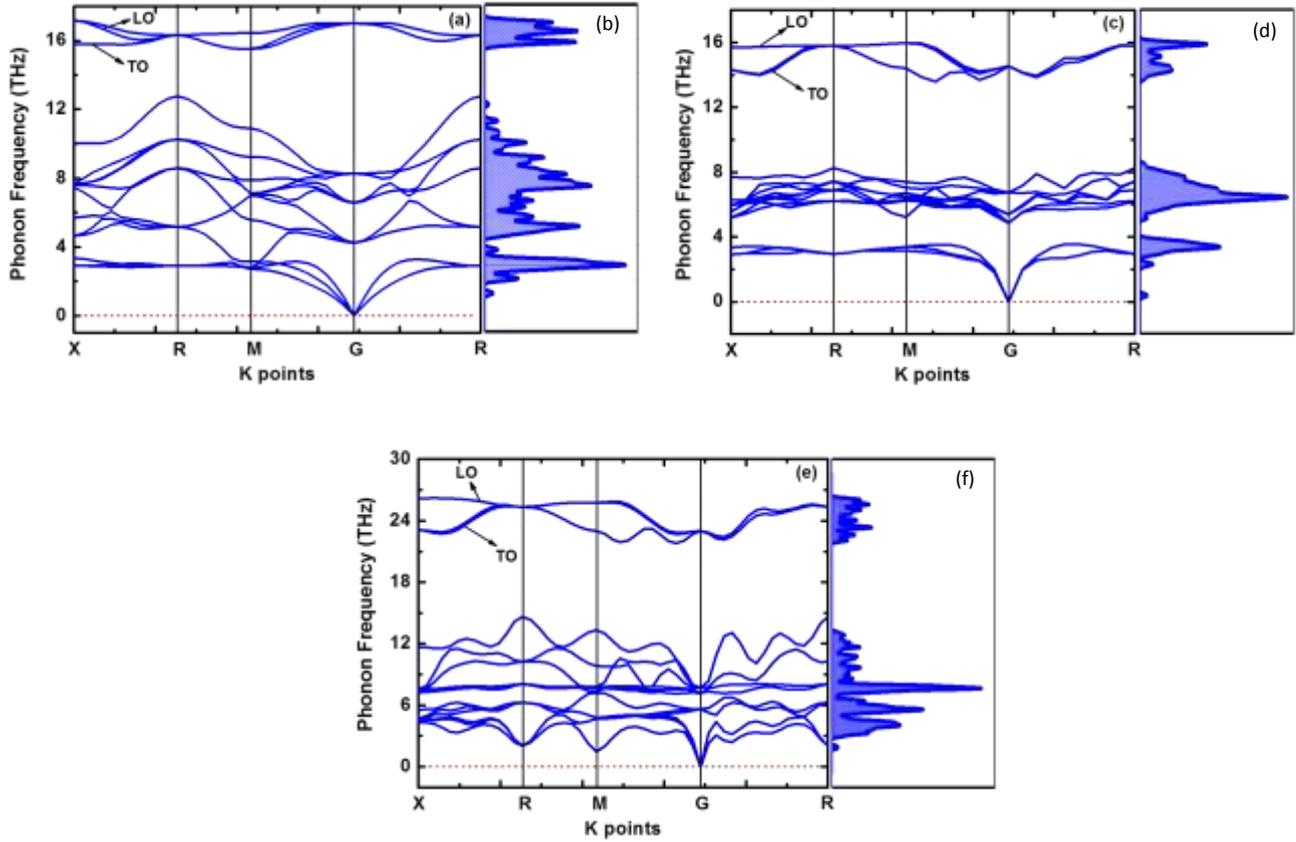

**Fig. 5.** The PDC and PHDOS of $Ti_3TlN$ (a, b), $Ni_3SnN$ (c, d) and $Co_3AlC$ (e, f) compounds. Zero phonon frequency is indicated by the red dashed line.

3.6 *Thermodynamical properties*

Temperature-dependent thermodynamical potential functions namely Helmholtz free energy $F$, enthalpy $E$, entropy $S$, and the phonon specific heat $C_v$ and the Debye temperature $\Theta_D$ at zero pressure have been calculated using quasi-harmonic approximation [58, 59] as shown in Figs. 6. The parameters are evaluated in the temperature range of 0-1000 K where harmonic model is assumed to be valid and no phase transitions are expected for the anti-perovskites considered here. The Helmholtz free energy (F) of the three phases is found to decrease gradually with increasing temperature as depicted in Fig. 6(a). This behavior is canonical since thermal disturbance adds to disorder and the entropy of the compounds increase. The difference between internal energy of a system and the amount of unused energy to perform work (represented by the product of entropy, S, of a system and the absolute temperature, T) is called the free energy. The increasing trend of the internal energy ($E$) with temperature is observed for the phases as expected for compounds in different phases (Fig. 6a). It is noted from Fig. 6 (a) that for the compounds considered, below 100 K, the values of $E$, $F$ and TS are almost zero.

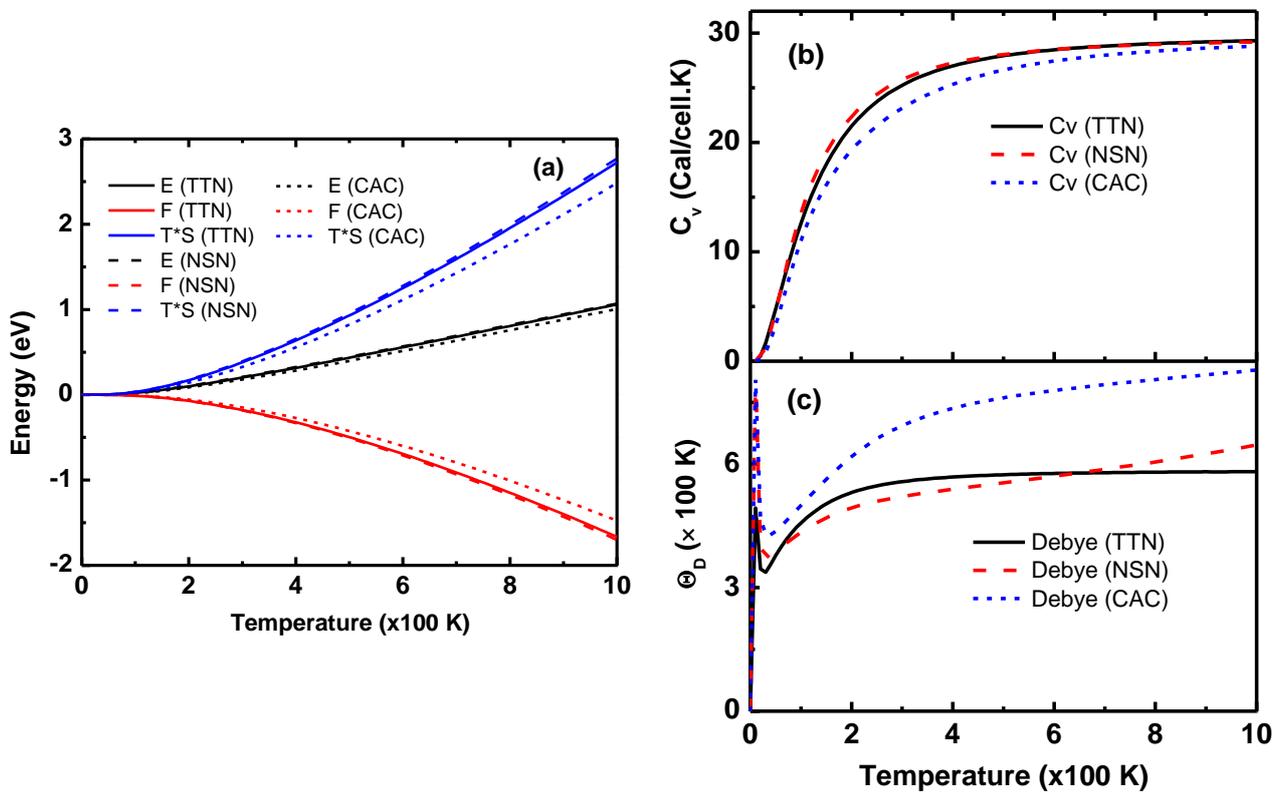

**Fig. 6.** The temperature dependence of the thermodynamical potential functions, $C_v$ and $\Theta_D$ of $Ti_3TlN$, $Ni_3SnN$ and $Co_3AlC$ compounds.

The specific heat at constant volume, $C_v$ of the anti-perovskites are calculated and illustrated in Fig. 6b. It is well known that phonon thermal softening happens with increasing temperature, consequently the heat capacity also increases with increasing temperature. The curves reveal that the trend of $C_v$ for the phases TTN, NSN and CAC are almost identical; however $C_v$ is a bit lower for the CAC phase. The overall trends of $C_v(T)$ can be explained by the Debye model. In the low temperature limit, $C_v$ exhibits the Debye-$T^3$ behavior [60]. At high temperature limit, the curves follow Dulong-Petit law where the $C_v$ does not depend strongly on the temperature [61].

The temperature dependence of the Debye temperature $\Theta_D$ has also been calculated using the PHDOS curves as illustrated in Fig. 6(c). The vibration frequency of particles is changed with temperature that is reflected by the Debye temperature, $\Theta_D$. Moreover, it is also linked to the bonding strength. The value of $\Theta_D$ is higher for the CAC compound than that of TTN and NSN which is in consistent with our previous stiffness based discussions in Section 3.2. The value of $\Theta_D$ is also estimated through the calculation of average sound velocity using the formalism described elsewhere [62] and are represented in Table 6.

**Table 6:** Calculated density ($\rho$ in gm/cm$^3$), longitudinal, transverse and average sound velocities ($v_l$, $v_t$, and $v_m$, respectively, all in km/s) and Debye temperature ($\theta_D$ in K) of Ti$_3$TlN, Ni$_3$SnN and Co$_3$AlC compounds.

| Compound | $\rho$ | $v_l$ | $v_t$ | $v_m$ | $\Theta_D$ |
|---|---|---|---|---|---|
| Ti$_3$TlN | 4.3 | 6.8 | 4.17 | 4.6 | 452 |
| Ni$_3$SnN | 3.7 | 7.3 | 4.52 | 4.98 | 489 |
| Co$_3$AlC | 1.46 | 1.16 | 7.16 | 7.89 | 775 |

## 3.7 Optical properties

Time dependent perturbations to the electronic ground states are used to define the interaction between the photons and valence and conduction electrons of a system. Electronic transitions occur between occupied states below the $E_F$ to unoccupied states above the $E_F$ due to the electromagnetic field that shines onto a material with sufficient energy. Therefore, the optical properties of the materials are mainly due to the interband and intra-band contributions with the later coming from the low energy infrared part of the spectra in metal and metal-like systems. An energy smearing of 0.5 eV has been used to specify the Gaussian broadening for all optical parameter calculations. The optical properties of studied compounds are computed from the complex dielectric function and a semi-empirical Drude term (with unscreened plasma frequency of 3 eV and damping of 0.05 eV) is also employed to calculate the dielectric function since the compounds are metallic in nature that have already been discussed in Section 3.3. The optical properties of the TTN, NSN and CAC have been evaluated for photon energies up to 30 eV for the electric field polarization vectors along [100] and [001] directions for the first time. We have found that the optical spectra for the two polarization directions are similar in nature and are almost identical. This is somewhat expected for cubic crystals. Therefore, data only for [100] direction has been shown in Figs. 7 (a-h).

The optical properties of TTN, NSN and CAC are determined using the frequency-dependent dielectric function $\varepsilon(\omega) = \varepsilon_1(\omega) + i\varepsilon_2(\omega)$. The imaginary part of dielectric function, $\varepsilon_2(\omega)$ is calculated from the transition matrix elements between the occupied and unoccupied electronic states as [62]

$$\varepsilon_2(\omega) = \frac{2e^2\pi}{\Omega\varepsilon_0} \sum_{k,v,c} \left|\psi_k^c | \bm{u.r} | \psi_k^v\right|^2 \delta\left(E_k^c - E_k^v - E\right)$$

Here $\omega$ is the light frequency, $\boldsymbol{u}$ is the vector defining the polarization of the incident electric field, $e$ is the electronic charge, $\psi_k^c$ and $\psi_k^v$ are the conduction and valence band wave functions at $k$, respectively. The real part of dielectric function ($\varepsilon_1$) can be evaluated using the Kramers-Kronig equation which relates the real part with the imaginary part.

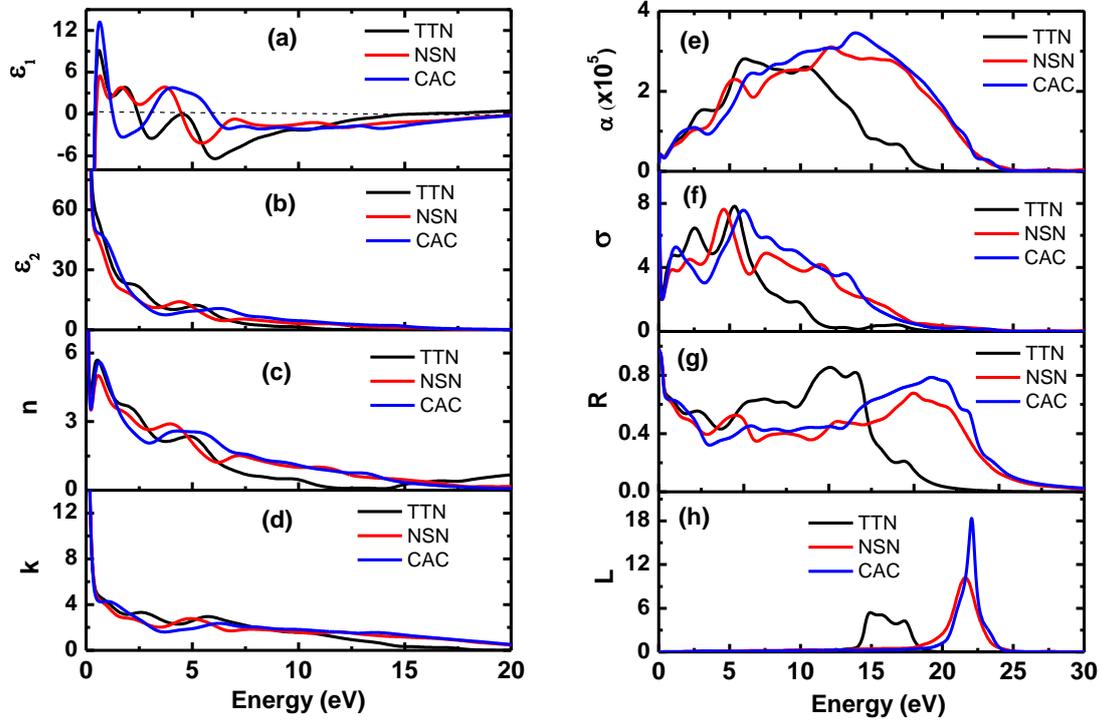

**Fig. 7.** Energy dependence of dielectric function (a) real part (b) imaginary part, (c) refractive index, (d) extinction coefficient, (e) absorption coefficient, (f) photo conductivity, (g) reflectivity, (d) loss function of the Ti$_3$TlN (TTN), Ni$_3$SnN (NSN), and Co$_3$AlC (CAC) compounds for [100] electric field polarization.

The real part of the dielectric function, $\varepsilon_1(\omega)$ for the phases TTN, NSN and CAC are illustrated in Fig. 7a. The intra-band transition of electrons has significant contribution to the optical properties of the materials and mainly affects the low energy region (infrared region) of the spectra. The low energy peaks are found at ~ 0.6, 0.65, and 0.63 eV in the $\varepsilon_1(\omega)$ curves for the TTN, NSN and CAC phases, respectively, due to the intraband transition of electrons.

The higher negative value $\varepsilon_1(\omega)$ implies a Drude-like behavior of the materials which is consistent with our discussion in Section 3.3. The imaginary part, $\varepsilon_2(\omega)$ of the compounds are depicted in Figs. 7b. The $\varepsilon_1(\omega)$ shows zero value from below at around 15.5, 21.5 and 22.3 eV (Fig. 7a) and the $\varepsilon_2(\omega)$ approaches zero from above at around 19.7, 25.8, and 24.8 eV (Fig. 7b) for the TTN, NSN and CAC compounds, respectively, which indicates further the metallic nature of the studied anti-perovskites.

The frequency dependent refractive index $n$ and the extinction coefficient $k$ have been presented in Fig. 7(c, d). The high value of $n$ of a material is an essential factor to design optoelectronic devices. The static value of $n(0)$ is found to be 8.01, 7.47 and 7.57 with the main peaks appearing at 0.53, 0.56 and 0.6 eV for the phases TTN, NSN and CAC, respectively. The absorption coefficient is related to the extinction coefficient by $k$ ($\alpha = 4\pi k/\lambda$). The sharp peaks are obtained at 2.6, 2.27 and 1.08 eV for TTN, NSN and CAC, respectively. These peaks are characterized by the intra-band transitions of electrons (Fig. 7d).

The energy loss of the light passing through materials can be measured by the absorption coefficient. The energy dependence of absorption coefficient, $\alpha$ have been evaluated and are shown in Fig. 7(e). The value of $\alpha$ is dominant in the ultraviolet (UV) region while it is weak in the infrared (IR) region. It is noted that the value of $\alpha$ increases in the direction of UV region and reaches a maximum value at 6.0, 12.2 and 13.9 eV for the TTN, NSN and CAC compounds, respectively. The highest absorption spectrum is found for the CAC material. The $\alpha$ is high in wide energy range that could be used in optoelectronic devices in both visible and ultraviolet regions. As expected, due to the metallic nature of studied compounds, the photoconductivity starts at zero photon energy as shown in Fig. 7f. The TTN compound exhibits the highest photo-conductivity at 5.4 eV, and for others the maxima showed up at 4.6 eV (NSN) and 6 eV (CAC).

The reflectivity (R) spectra as a function of energy for the studied phases are illustrated in Fig. 7g. It is seen that the value of R starts with a value of 97% for all phases and rises to maximum values of 85% at 12.0 eV, 68% at 18.0 eV and 79% at 19.27 eV for the TTN, NSN, and CAC compounds, respectively. It is reported that a compound with reflectivity ~ 44% in the visible light region would be capable of reducing solar heating [63]. It is noted that the reflectivity is always above 44% in case of TTN compound. For the phases NSN and CAC, the R values lie below 40% in the IR (1.24 meV - 1.7 eV) and visible region (1.7 eV – 3.3 eV), while in the UV region, it is above 44%. Therefore, the TTN ($Ti_3TlN$) compound is a potential candidate for the practical use as a coating material to reduce solar heating. The reflectivity spectra for all phases approach to zero at the incident photon energy range of 24 to 30 eV.

The important feature of loss spectrum can be understood using the dielectric function that defines energy loss of an electron passing through the material. There is no loss observed up to 13 eV for all the studied compounds. The loss peaks appeared in the energy range from 13 to 25 eV. The peaks are prominent in the energy range where the value of $\varepsilon_1 = 0$ and $\varepsilon_2 < 1$. The loss peaks are associated with the so called bulk plasma frequency $\omega_P$ that are found to be at 14.9, 21.6 and 22 eV for the TTN, NSN and CAC compounds, respectively. The $\omega_P$ also marks the onset of the fast drop of the reflectivity (R) of the materials under consideration, as shown in Fig. 7g.

## 4. Conclusions

First principles calculations based on the density functional theory (DFT) has been employed to study thermodynamic parameters, optical properties, Fermi surface topology, Mulliken bond overlap population and Vickers hardness of recently reported damage tolerant $Ti_3TlN$ (TTN), $Ni_3SnN$ (NSN) and high stiffness $Co_3AlC$ (CAC) anti-perovskite materials for the first time. We have revisited the structural, elastic and electronic properties of the compounds to emphasis the reliability of our calculations. The traditional mechanical stability conditions have been confirmed by the evaluated elastic constants $C_{ij}$ of the phases under consideration. The degree of anisotropy of the compounds is evaluated by the distortion from the spherical shape of the elastic moduli using the contour plot (3D) and 2D projections as well as by computed anisotropic index. The conduction and valence bands are overlapping at the $E_F$ consequently no band gap has been observed indicating metallic nature of the compounds. N-$2p$ and Tl-$6s$, N-$2s$ and Sn-$5s$, C-$2p$ and Al-$3p$ orbital states and Ti-$3d$, Ni-$3d$ and Co-$3d$ electronic states for the phases TTN, NSN and (CAC), respectively, are mostly contributing to the TDOS at the $E_F$. The Pugh and Poisson's ratios affirm that the compounds are ductile in nature. The strong metallic-covalent bonding is formed between Co-Al, and Ni-Sn atoms in the CAC and NSN phases, respectively, based on the strong charge accumulation. On the other hand, robust metallic bond is detected between Ti and Tl in the TTN phase due to the absence of charge accumulation that are related to further strengthening of the damage tolerant properties. The Vickers hardness values are found to be 3.6, 3.58 and 4.04 GPa for TTN, NSN and CAC, respectively, indicating the moderately hard nature and easy machinability of the studied compounds by conventional cutting tools. No negative frequency is observed in the phonon dispersion curves; consequently the studied compounds are predicted to be dynamically stable. The Debye temperatures are calculated as 452 K, 489 K and 775 K for the phases TTN, NSN and CAC, respectively. The static refractive indices $n(0)$ are found to be 8.01 (TTN), 7.47 (NSN) and 7.57 (CAC) that suggest the materials could be used to design optoelectronics appliances. The value of reflectivity for the TTN compound up to mid-UV region is always above 44%. This makes the compound a potential candidate for the practical use as a coating material to minimize solar heating.

# Acknowledgements

Authors are grateful to the Department of Physics, Chittagong University of Engineering & Technology (CUET), Chattogram-4349, Bangladesh, for providing the financial support for this work.